\documentclass[3p,fleqn,preprint,sort&compress]{elsarticle}
\usepackage{amsmath,amssymb,natbib}
\usepackage[english]{babel}
\usepackage[T2A]{fontenc}
\usepackage{epsfig}
\usepackage[utf8]{inputenc}
\usepackage{graphicx}
\usepackage{bm}
\usepackage[unicode,colorlinks=true,urlcolor=blue,citecolor=blue,linkcolor=blue]{hyperref}
\usepackage{lineno}

\begin{document}
\title{Interplay of magnetism and superconductivity in 2D extended Hubbard model}

\author{V.\,F. Gilmutdinov\corref{cor1}}
\author{M.\,A. Timirgazin}
\author{A.\,K. Arzhnikov}

\cortext[cor1]{vital@udman.ru}
\address{{UdmFRC UrB RAS},
	{T. Baramzina st., 34},
	{Izhevsk},
	{426067},
	{Udmurt republic},
	{Russia}}

\journal{Journal of Magnetism and Magnetic Materials}

\begin{highlights}
	\item Interplay between magnetic and superconducting orders is studied within extended Hubbard model on a square lattice. The symmetry of ordered states is considered in a broad form, spiral magnetic states and intermediate $s$+i$d$-wave superconductivity being taken into account.
	\item Both macroscopic phase separation and microscopic coexistence between magnetic and superconducting phases are considered.
	\item Ground state phase diagrams of the model are constructed in variables of band filling and interactions magnitudes.
	\item Coexistence of spin spiral and superconducting states is found in a wide range of model parameters. Phase separation between antiferromagnetic insulator and superconducting state is realized near half-filling.
	\item Mutual suppression of magnetization and superconducting order parameter is observed in coexistence regime. Dome-shaped dependence of superconducting order parameter on doping is observed.
\end{highlights}

\begin{abstract} 
	The interplay between magnetic and superconducting states on a square lattice is studied using the extended Hubbard model, which takes into account the attraction of electrons located at nearest neighbor sites. Ferro-, antiferro-, and spiral magnetic states with all possible sets of spiral wave vectors, as well as singlet superconducting states with $s$-wave and $d$-wave pairing order parameters, are considered. Formation of a state with the intermediate $ s + $i$d $-symmetry and phase separation between different phases are allowed. The results of the Hartree-Fock and slave boson approximations are compared in order to study the role of correlation effects.
Both macroscopic phase separation and microscopic coexistence of the superconducting and magnetic phases are found under certain model parameters.	
\end{abstract}


\maketitle 
\section{Introduction}\label{intro}

In recent decades, significant advances have been made in the study of the magnetic and superconducting properties of high-temperature superconductors (HTSC). Competition and the mutual influence of magnetism and superconductivity are being intensively studied. 
For a number of compounds 
based on copper oxides and iron pnictides (and others) 
 both coexistence of superconductivity and commensurate or incommensurate magnetic orders and separation into magnetic and superconducting phases are found~\cite {Sidis01, Yu07,Takeshita09,Lee99, Miller06,Kibune22}. 

For the quasi-two-dimensional superconducting compounds the system instability in respect to  the formation of incommensurate and non-collinear spin density wave (SDW) ordering is established. The question of which type of the superconducting order parameter is realized in both copper oxides and iron pnictides is still open. Various symmetries are studied in the literature: singlet $s$-~\cite{Tsuei00, Mazin08} and $ d $-wave states~\cite{Huang90, Maier11}, intermediate states like $ s + $i$d $~\cite {Ruckenstein87, Kotliar88}, $ d + $i$d $~\cite {Laughlin98, Kreisel17}, etc. 

In the framework of Hubbard model~\cite{Hubbard63}, which is traditionally applied to $ 3d $-metals compounds, an interplay between magnetism and superconductivity is revealed. Using the random phase approximation Scalapino et al. show that the SDW type magnetic ordering, arising from the Fermi surface nesting, leads to conditions favorable for $ s $- and $ d $-wave superconductivity, and the system is sensitive to the band structure and its filling~\cite {Scalapino86, Scalapino87}.
The spin-fluctuation mechanism of the Cooper pairing is studied by the authors of Ref.~\cite {Romer15} in the weak-coupling limit of the Hubbard model at $ T = 0$. The phase diagrams constructed in this work include a rich variety of singlet and triplet superconducting states, and the $ d $-wave symmetry of the superconducting order parameter remains the ground state near the half-filling.

The competition between antiferromagnetic (AF) order and $ d $-wave superconductivity is considered in a number of investigations within the Hartree-Fock approximation (HFA) for the Hubbard model. Both the microscopic coexistence and the macroscopic phase separation (PS) of the states are found~\cite {Ghosh99,Reiss07}. The authors of~\cite {Ghosh99} take into account the $ s $- and $ s + d $-wave pairings, emphasizing that in their model the magnetic state is more stable than superconducting one due to the Fermi surface nesting. The coexistence and PS between AF and superconducting states are shown within the framework of the Monte Carlo method~\cite{Kobayashi10,Yokoyama13} and the renormalization group functional~\cite {Reiss07, Yamase16}, as well as between $ s $-wave superconductivity and AF (commensurate and incommensurate) ordering in the weak-coupling limit~\cite {Vorontsov09, Fernandes10}.

In the slave boson approach (SBA) the extended $s$-wave pairing is shown to be stabilized for a limited doping range ~\cite{Kopp88}. 
A significant difference in the results of the HFA and SBA approaches 
in vicinity of the half-filling for $U/W\approx 1$ ($W$ is the bandwidth) is found~\cite{Bulka96}. In the weak-coupling limit, the gap in the excitation spectrum at $T = 0$ decreases for SBA in comparison with the value obtained in HFA. The difference between the two approaches is maximal near half-filling and decreases near the band edges. Qualitative and quantitative corrections of SBA to HFA was shown: the energy gap agrees with the HFA only in the small-density limit~\cite{Bak98}. 
The results of SBA method show that electronic correlations significantly change the properties of the superconducting phase~\cite{Bulka98}.
The appearance of superconductivity with extended $ s $- and $ d $-wave symmetries of the superconducting order parameter for the heavy fermion systems is considered in the paper \cite{Sacramento10} as a function of Coulomb repulsion. The calculation results show that, if the attractive interaction is not too weak, superconductivity is retained with an increase in $ U $ and prevails for all band fillings. Superconductivity is suppressed at high $ U $ only near half-filling (in particular, for $ d $-wave symmetry).

Despite the large number of studies presented in the literature, all of them appear to be limited in one way or another: all possible superconducting or non-collinear magnetic states are not taken into account; the studies themselves are performed in the selected areas of the model parameters; the approximations used do not take into account electronic correlations; a possibility of the coexistence or PS between the magnetism and superconductivity is ignored.
A study that systematically considers the competition between superconductivity with the mixed order parameter symmetry and spiral magnetic states on a square lattice, and establishes the role of electronic correlations, is not performed.

The conditions for the formation of spiral magnetic states using the HFA and SBA approximations in the Hubbard model on square and cubic lattices are studied in~\cite {Igoshev09, Igoshev13}. The phase diagrams of the model in terms of the Hubbard repulsion $U$ 
and the band filling $ n $ include a variety of spiral magnetic phases, as well as PS between them. Comparison of HFA and SBA results show that electron correlations significantly suppress the magnetic states. 

In this paper, we present the results of a study of the two-dimensional single-band extended Hubbard model within HFA and SBA approaches. 

\section{Formalism}
We study the mutual influence of magnetism and superconductivity using the Hubbard model extended by a term describing the attraction of electrons located at nearest neighbor sites, $\hat{V}$:
\begin{equation}\label{coex_ham}
	\begin{array}{l}
		\displaystyle \hat{H} =\hat{K} + \hat{U} - \hat{V},\\[15pt]
		\displaystyle \hat{K} = \sum_{j,j',\sigma} t_{j,j'}c^\dag_{j,\sigma}c_{j',\sigma}-\mu\sum_{j,\sigma}c^\dag_{j,\sigma}c_{j,\sigma},\\[15pt]
		\displaystyle \hat{U} = U\sum_{j}n_{j,\uparrow}n_{j,\downarrow}= U\sum_{j}c^\dag_{j,\uparrow}c_{j,\uparrow}c^\dag_{j,\downarrow}c_{j,\downarrow},\\[13 pt]
		\displaystyle \hat{V} = V_0 \sum_{j,j'}n_{j,\uparrow}n_{j',\downarrow}=V_0 \sum_{j,j'}c^\dag_{j,\uparrow}c^\dag_{j',\downarrow}c_{j',\downarrow}c_{j,\uparrow},
	\end{array}
\end{equation}
where $t_{j, j'}$ is the matrix of electron transfer integrals (we take into account the nearest and next-nearest neighbor sites with integrals $-t $ and $t'$, respectively), $c^\dag_{j,\sigma}$ and $c_{j',\sigma} $ are the creation and annihilation operators of electrons at a site $j$ with spin $\sigma$, $U$ is the on-site Coulomb repulsion parameter, $V_0$ is an attraction parameter between nearest neighbor sites, which is responsible for the formation of Cooper pairing, $\mu$ is the chemical potential, $n_{j, \sigma} = c^\dag_{j,\sigma} c_{j,\sigma}$ is the operator of electron number at the site $j$ with spin $\sigma$.

So far, no consensus has been reached on the nature of HTSC. We use the attraction term $\hat V$ to get peculiarities of superconducting state, the mechanism of the electron attraction being not specified, but we assume it to be driven by either AF spin fluctuations~\cite{Izyumov99}, or the resonating valence bond mechanism~\cite{Kotliar88}.
The Hamiltonian with the effective attraction between different site fermions was used to describe the $d$-wave pairing~\cite{Mayr05} and allows one to yield the superconducting state of $s$-wave, $d$-wave and intermediate $s+$i$d$-wave type~\cite{Timirgazin19,Micnas02}.

Assuming the inter-site $\hat{V}$ interaction to be weaker than the on-site $\hat{U}$, we make the mean field approximation for the first one:
\begin{equation}
\displaystyle {\hat V} = \dfrac{1}{2}\sum_{j,j'}\left( \Delta_0\exp{(i\phi_{j,j'})}c^\dag_{j,\uparrow}c^\dag_{j', \downarrow}+h.c. \right)-\dfrac{N|\Delta_0|^2}{V_0},
\end{equation}
where $N$ is the number of lattice sites.
Here the order parameter is introduced: $V_0\langle c^\dag_{j,\uparrow}c^\dag_{j',\downarrow} \rangle \equiv \Delta_0{\exp(-i\phi_{j,j'})}/2$, where the phase shift  $\phi_{j,j'}$ is homogeneous and depends on the mutual arrangement of sites $j$ and $j'$.

In order to treat spiral magnetic order the local spin rotation by the angle $\mathbf{QR}_j$ is applied to the Hamiltonian, where $\mathbf{Q}$ is a spiral wave vector. This makes mapping to an effective ferromagnetic state with non-diagonal hopping and superconducting terms: $t_{j,j'}\rightarrow t^{\sigma,\sigma'}_{j,j'}$, $\Delta_0\rightarrow \Delta^{\sigma,\sigma'}_{j,j'}$~\cite{Igoshev15}. In the Kotliar and Ruckenstein formulation of SBA~\cite{Kotliar86} bosonic operators $e_j$, $p_{j,\sigma} $ and $d_{j}$ are introduced, corresponding to empty, once and twice occupied sites $j$, and constrains are imposed that exclude nonphysical states:
\begin{equation}
	\begin{array}{l}
		\displaystyle e^\dag_j e_j+\sum_{\sigma}p^\dag_{j,\sigma}p_{j,\sigma}+d^\dag_j d_j=1,\\
		\displaystyle p^\dag_{j,\sigma}p_{j,\sigma}+d^\dag_j d_j=c^\dag_{j,\sigma}c_{j,\sigma}.\\
	\end{array}
\end{equation}
The replacement $c_{j,\sigma} \rightarrow z_{i,\sigma} c_ {j,\sigma}$ ensures the coherence of bosonic and fermionic fields. In the introduced parametrization, the Hamiltonian takes the  diagonal form with respect to the bosonic operators:
\begin{equation}\label{ham_sba_node}
	\begin{array}{c}
		\displaystyle {\cal H} = \sum_{j,j',\sigma,\sigma'}t^{\sigma,\sigma'}_{j,j'} c^\dag_{j,\sigma}c_{j',\sigma'} z^\dag_{j,\sigma}z_{j',\sigma'}+U\sum_j d^\dag_j d_j +
		\\[13pt] \displaystyle + \dfrac{1}{2}\sum_{j,j',\sigma,\sigma'} \Delta_{j,j'}^{\sigma,\sigma'}c^\dag_{j,\sigma}c^\dag_{j',\sigma'}z^\dag_{j,\sigma}z^\dag_{j',\sigma'}+h.c.-\dfrac{N|\Delta_0|^2} {V_0}.
	\end{array}
\end{equation} 
Using the static and saddle point approximations, the thermodynamic potential of the grand canonical ensemble of the system can be written as~\cite{Igoshev15}:
\begin{equation}\label{ensemble}
	\begin{array}{c}
\displaystyle \Omega=\eta\left(e^2+p_\uparrow^2+p_\downarrow^2+d^2-1\right)+Ud^2-\\[13pt]
\displaystyle - \sum_\sigma \lambda_\sigma\left(p_\sigma^2+d^2\right)+\dfrac{\Delta_0^2}{V_0}+\Omega_f ,
\end{array}
\end{equation}
where $\lambda_\sigma$ and $\eta$ are Lagrange multipliers. The fermionic part $\Omega_f$ of the potential (\ref{ensemble}) after the Fourier transform can be represented in the matrix form:
\begin{equation}
\Omega_f = \dfrac{1}{2}\sum_\mathbf{k} {\hat \gamma}^\dag_\mathbf{k}{\hat{\cal T}}_\mathbf{k}^f{\hat \gamma}^{}_\mathbf{k},
\end{equation}
where  ${\hat \gamma^\dag_\mathbf{k}}=\begin{pmatrix} c^\dag_{\mathbf{k-Q}/2,\uparrow} & c_{\mathbf{-k+Q}/2,\downarrow} & c^\dag_{\mathbf{k+Q}/2,\downarrow}  & c_{\mathbf{-k-Q}/2,\uparrow} \end{pmatrix}$, $\hat{\cal T}_\mathbf{k}^f$ is $4\times 4$ square matrix:
\begin{equation}\label{matrix}
	{\hat {\cal T}}_\mathbf{k}^f=\begin{pmatrix}
		z^2_\uparrow \varepsilon_{\mathbf{k},+}-\mu+\lambda_\uparrow & -z_\uparrow z_\downarrow \Delta_\mathbf{k,+} & z_\uparrow z_\downarrow \varepsilon_\mathbf{k,-} & z^2_\uparrow \Delta_\mathbf{k,-}\\
		-z_\uparrow z_\downarrow \Delta_{\mathbf{k},+}^* & -\left(z^2_\downarrow \varepsilon_\mathbf{k,+}-\mu+\lambda_\downarrow\right) & -z_\downarrow^2 \Delta_{\mathbf{k},-}^* & z_\uparrow z_\downarrow \varepsilon_{\mathbf{k},-} \\
		z_\uparrow z_\downarrow \varepsilon_\mathbf{k,-} & -z^2_\downarrow \Delta_\mathbf{k,-} & z^2_\downarrow \varepsilon_\mathbf{k,+} -\mu+\lambda_\downarrow & z_\uparrow z_\downarrow \Delta_\mathbf{k,+} \\
		z^2_\uparrow \Delta_\mathbf{k,-}^* & z_\uparrow z_\downarrow \varepsilon_\mathbf{k,-} & z_\uparrow z_\downarrow \Delta_\mathbf{k,+}^* & -\left( z^2_\uparrow \varepsilon_\mathbf{k,+}-\mu+\lambda_\uparrow \right) \\
	\end{pmatrix}
\end{equation}
Here
\begin{equation} 
	\begin{array}{l}
	\varepsilon_{\mathbf{k},\pm}=\left(\varepsilon^0_{\mathbf{k}+\mathbf{Q}/2}\pm \varepsilon^0_{\mathbf{k}-\mathbf{Q}/2}\right)/2,\\[13 pt]  
	\varepsilon^0_\mathbf{k}=-2t(\cos{k_x}+\cos{k_y})+4t'\cos{k_x}\cos{k_y}, \\[13 pt]
	\Delta_{\mathbf{k},\pm}=\left(\Delta_{\mathbf{k}+\mathbf{Q}/2}\pm \Delta_{\mathbf{k}-\mathbf{Q}/2}\right)/2,\\ [13 pt]
	\Delta_\mathbf{k}=\frac{1}{2}\Delta_0\sum_{j,j'} \exp(i\phi_{j,j'})\exp\left( i\mathbf{k}\left( \mathbf{R}_j-\mathbf{R}_{j'} \right) \right), 
	\end{array}
\end{equation}
$\varepsilon_\mathbf{k}^0$ being the square lattice dispersion law. Choosing the phase shift in the form
\begin{equation}
	\phi_{j,j'} = 
	\begin{cases}
		\displaystyle \;\;\; \pi\alpha,\;  \mathbf{R}_j-\mathbf{R}_{j'} = (\pm 1,0),\\ 
		\displaystyle -\pi\alpha,\; \mathbf{R}_j-\mathbf{R}_{j'} = (0,\pm 1),
	\end{cases}
\end{equation}
we obtain an intermediate $s+$i$d$-wave superconducting order parameter:
\begin{equation}\label{result}
	\begin{array}{l}
		\displaystyle \Delta_\mathbf{k} = \Delta^s_\mathbf{k}\cos{\pi\alpha}+i\Delta^d_\mathbf{k}\sin{\pi\alpha}.
	\end{array}
\end{equation}
where:
\begin{equation}	
\Delta_\mathbf{k}^s=\Delta_0\left(\cos{k_x}+\cos{k_y}\right)
\end{equation}
is $s$-wave (extended $s$ or $s_{x^2+y^2}$) and
\begin{equation}
\Delta_\mathbf{k}^d=\Delta_0\left(\cos{k_x}-\cos{k_y}\right)
\end{equation}
is $d$-wave ($d_{x^2-y^2}$).
Varying $\alpha$ from $0$ to $\pi/2$ allows for a continuous transition from $s$- to $d$-wave pairing symmetry.

The quantum mechanical averaging of the fermionic part $\Omega_f$ of the thermodynamic potential (\ref{ensemble}) over the ground state of the Hamiltonian leads to the following result:
\begin{equation}
		\langle \Omega_f \rangle = \dfrac{1}{2}\sum_\mathbf{k}\langle {\hat \gamma}^\dag_\mathbf{k}{\hat{\cal T}}_\mathbf{k}^f{\hat \gamma}_\mathbf{k}  \rangle=\dfrac{1}{2}\sum_{\mathbf{k}}\left(E^{(1)}_{\mathbf{k}}+E^{(2)}_{\mathbf{k}}\right),
\end{equation}
where $E^{(1)}_{\mathbf{k}}$ and $E^{(2)}_{\mathbf{k}}$ are the negative spectrum branches, which must be determined numerically at each $\mathbf{k}$-point. Note that in HFA, where only the fermionic part is kept, the spectrum can be written explicitly:
\begin{equation}
	\begin{array}{c}
	\displaystyle E_{\mathbf{k}}=\pm\sqrt{(Um/2)^2+
		\varepsilon_\mathbf{k,+}^2+\varepsilon_\mathbf{k,-}^2+\Delta_\mathbf{k,+}^2+\Delta_\mathbf{k,-}^2\pm D_\mathbf{k}
	},\\[13 pt]
D_\mathbf{k}=2\sqrt{(\varepsilon_\mathbf{k,+}\varepsilon_\mathbf{k,-}+\Delta_\mathbf{k,+}\Delta_\mathbf{k,-})^2+(Um/2)^2(\varepsilon_\mathbf{k,+}^2+\Delta_\mathbf{k,+}^2)}.
\end{array}
\end{equation}
By numerically calculating the eigenvectors, the average values $n$, $m$ and $\Delta_0$ can be determined. Minimizing the thermodynamic potential (\ref{ensemble}) in respect to all the magnetic ($\mathbf{Q}$) and superconducting ($\alpha$) states at fixed parameters $U$, $V_0$, $t'$, $n$, one can construct the ground state phase diagrams of the system. 

\section{Results}
\begin{figure}
	{a)\includegraphics[width=\linewidth]{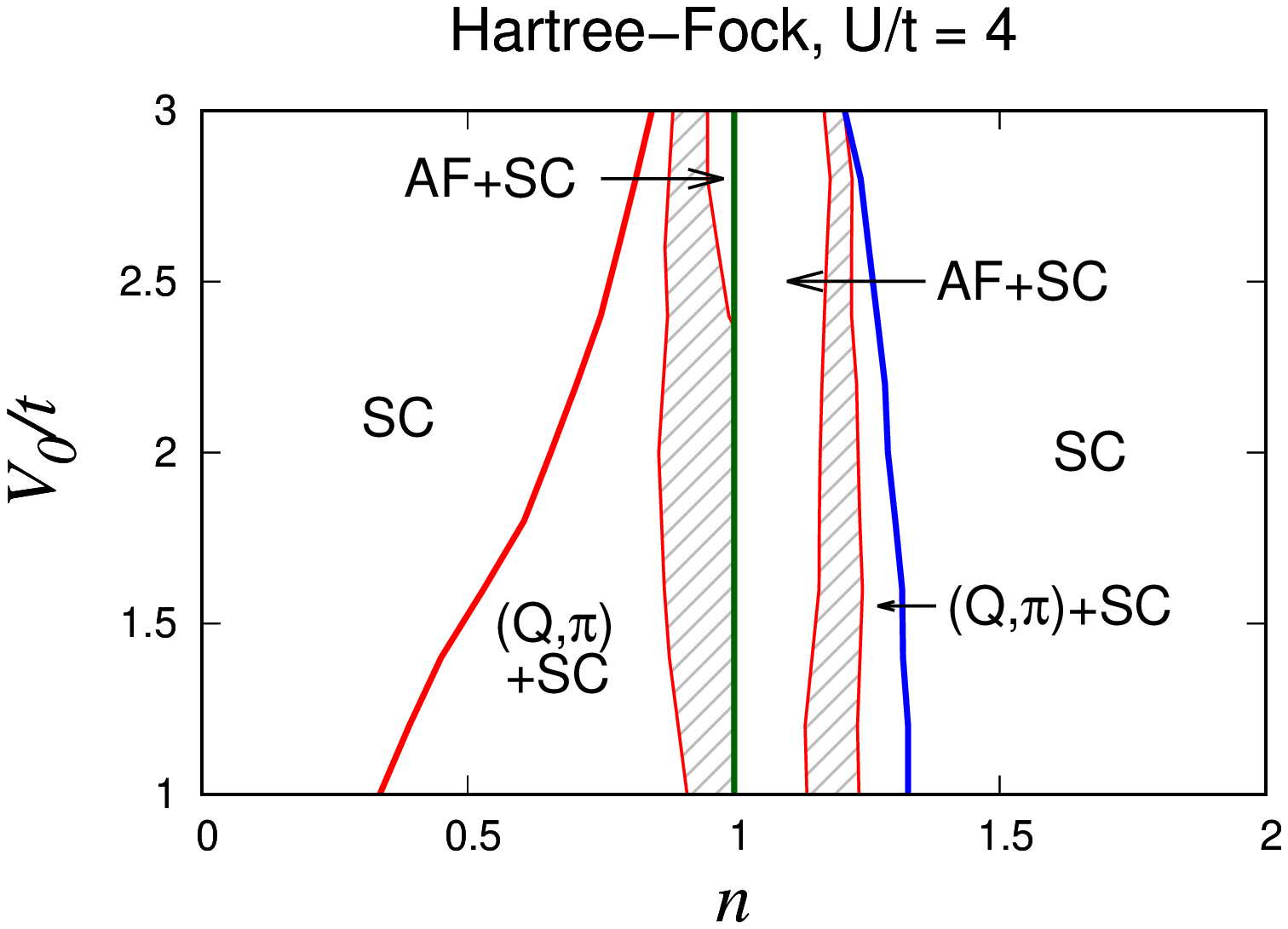}}
	{b)\includegraphics[width=\linewidth]{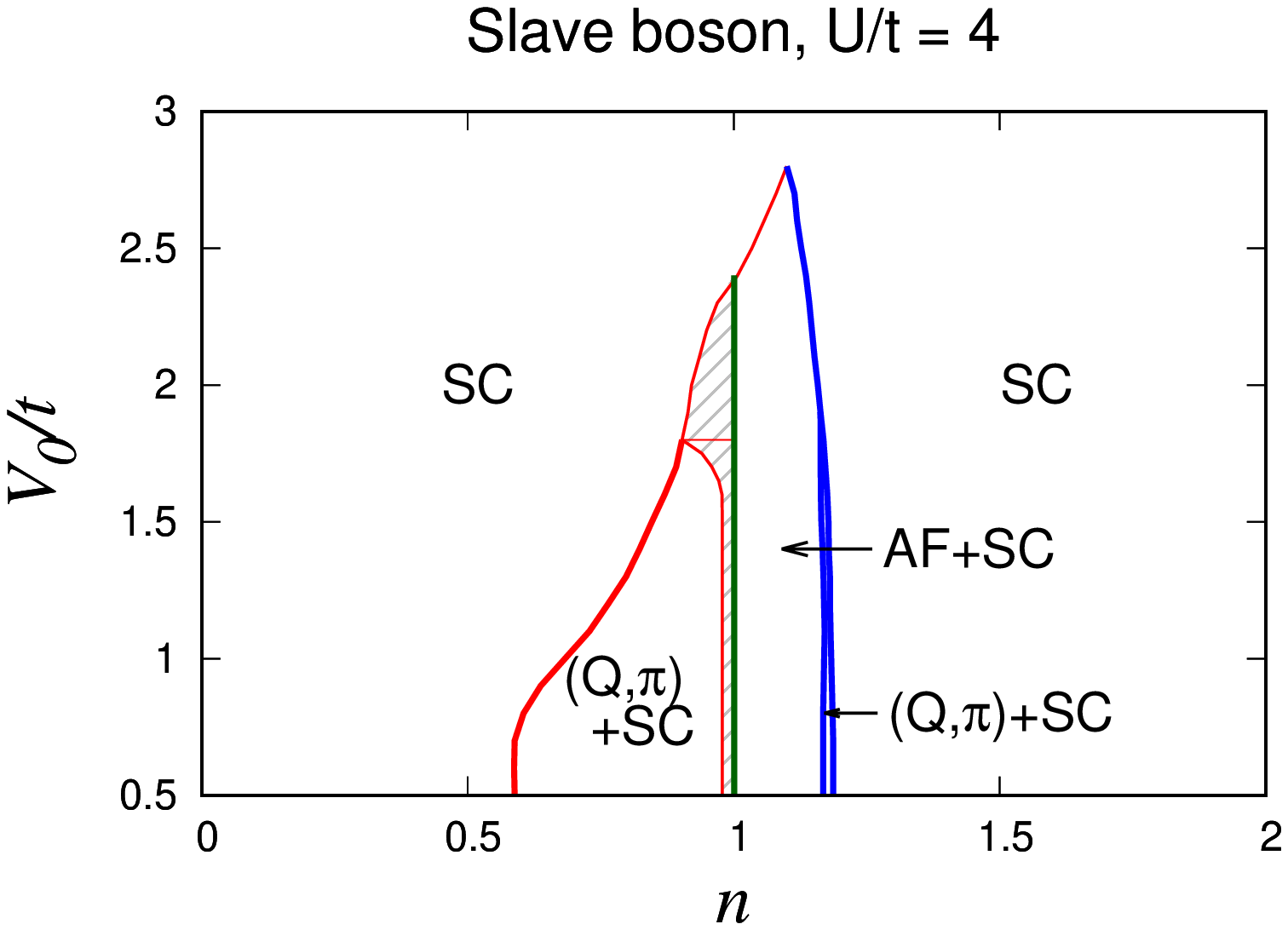}}
	\caption{Phase diagrams for $ U = 4t, t ' = 0.2t $: a) HFA, b) SBA. Thick blue lines --- the second order phase transitions, thick red lines --- the first order phase transition (narrow PS areas), thin red lines --- the boundaries of the PS areas (shaded). "SC" --- superconducting state,  " + " --- the coexistence of the magnetic and superconducting orders,   $(Q_1,Q_2)$ is the wave vector of the magnetic spiral. Thick green line $n=1$ --- AF insulating state. }
	\label{coex_4}
\end{figure}
\begin{figure}[h!]
	{a)\includegraphics[width=\linewidth]{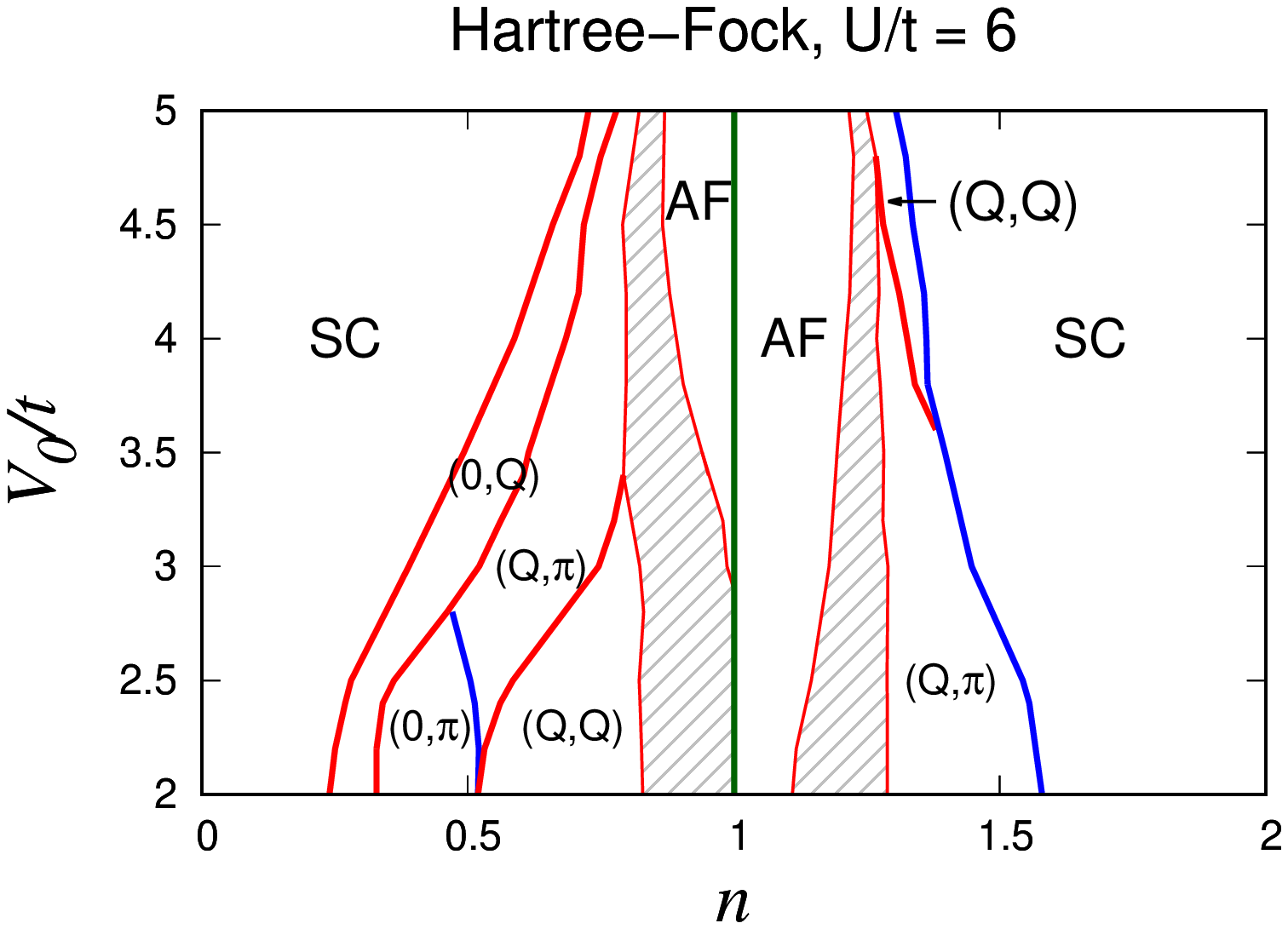}}
	{b)\includegraphics[width=\linewidth]{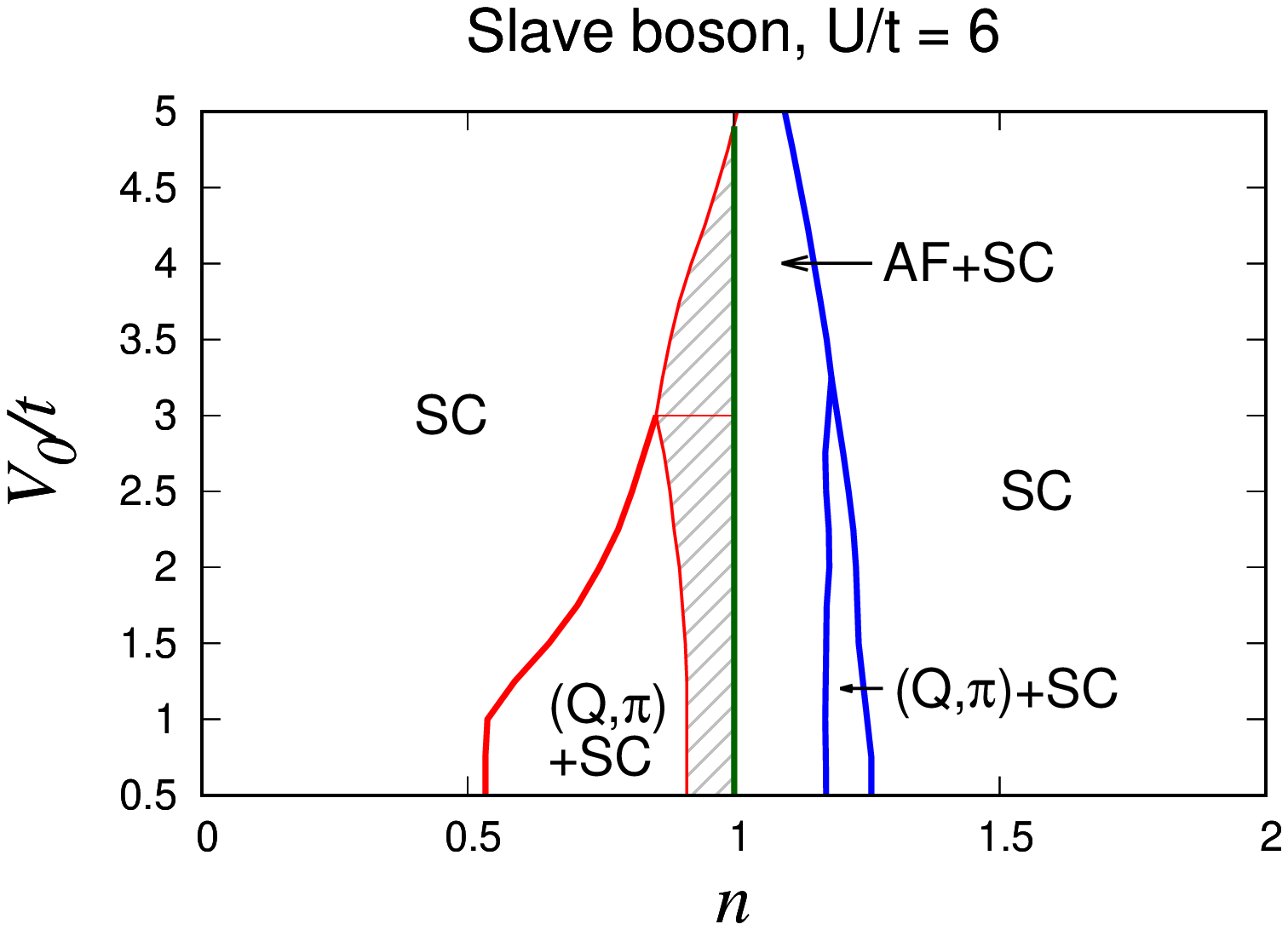}}
	\caption{Phase diagrams for $ U = 6t, t '/ t = 0 . 2 $.
		Notation is the same as in the figure \ref{coex_4}, but for a) index "SC" is omitted for all magnetic areas, the coexistence of magnetism and superconductivity is still being implied.}
	\label{coex_6}
\end{figure}
The results of our study consist of two parts: the phase diagrams of the model ground state are constructed in the variables of the superconducting electron attraction $V_0/t$ and the site electron concentration $n$ for the fixed values of Coulomb repulsion $U = 4t$ and $U=6t$; and the dependencies of the amplitudes of the magnetic moment $m$ and the superconducting gap $\Delta_0$ on the electron density $n$ for the values $V_0=1.5t$ and $U=4t$. The chosen values of $U$ and $t'$ correspond to HTSC based on copper oxides~\cite{Kuchinskii12,Hybertsen92}.  

\subsection*{1. $(V_0/t,n)$ phase diagrams}
To construct the phase diagrams, the ground state of the system should be determined on a grid of parameters $\mu$ and $V_0/t$ with fixed Coulomb repulsion $U$. For each set of parameters $(\mu,U,V_0,t'/t)$, the energy of all possible magnetic and superconducting states is calculated. The energies are then compared, and the state with the lowest energy is considered as the ground state. 

In the system, we detect phase transitions of the first and second order, and PS areas.
The PS areas boundaries are determined by two values of the electron density $n_1$ and $n_2$ corresponding to the same value of the Fermi level $\mu$. If the electron concentration is within this region, then two spatially separated phases are simultaneously realized in the system. We find the regions of the pure SC (magnetic order is absent), coexistence and pure AF (superconductivity is absent) insulating states in the system. Magnetic order have a spin-spiral structure with wave vector $(Q_1,Q_2)$.

The phase diagrams for $ U = 4t $ and  $ U = 6t $ in the cases of HFA and SBA are shown of the figures~\ref{coex_4} and \ref{coex_6}, correspondingly. Superconductivity is realized in the entire range of parameters under consideration with the only exception, which is the half-filling line, since the Fermi level lies in the energy gap, and the system becomes an AF insulator. 

It was found earlier that $s$-wave superconductivity is realized at low concentrations of charge carriers and there is $d$-wave superconductivity when approaching half filling, the transition between $s$-wave and $d$-wave states occuring through an intermediate $s+$i$d$-wave region~\cite{Timirgazin19,Micnas02}. The "{SC}"  area on the phase diagrams~\ref{coex_4}, \ref{coex_6} contains all above mentioned superconducting states, but in the region of existence of magnetic order the only $d$-wave state is realized.

Increasing of $U/t$ from 4 to 6 within HFA initiates the appearance of $(0,Q)$ and $(Q,Q)$ spin-spiral phases, which is in agreement with results of \cite{Igoshev15}.

Accounting for electron correlations in the SBA approach leads to the suppression of magnetic ordering, narrowing of the PS areas, and an expansion of the superconducting state region (Fig. \ref{coex_4},b)). It should be noted that magnetic order vanishes when increasing of $V_0/t$ due to the superconducting gap becomes greater than the AF one even for $n=1$.

In general, comparison of the diagrams for the HFA and SBA methods allows one to conclude that taking into account electronic correlations leads to the suppression of the range and variety of spiral magnetic states and the expansion of the superconductivity region. 
\begin{figure}[h]
	{a)\includegraphics[width=\linewidth]{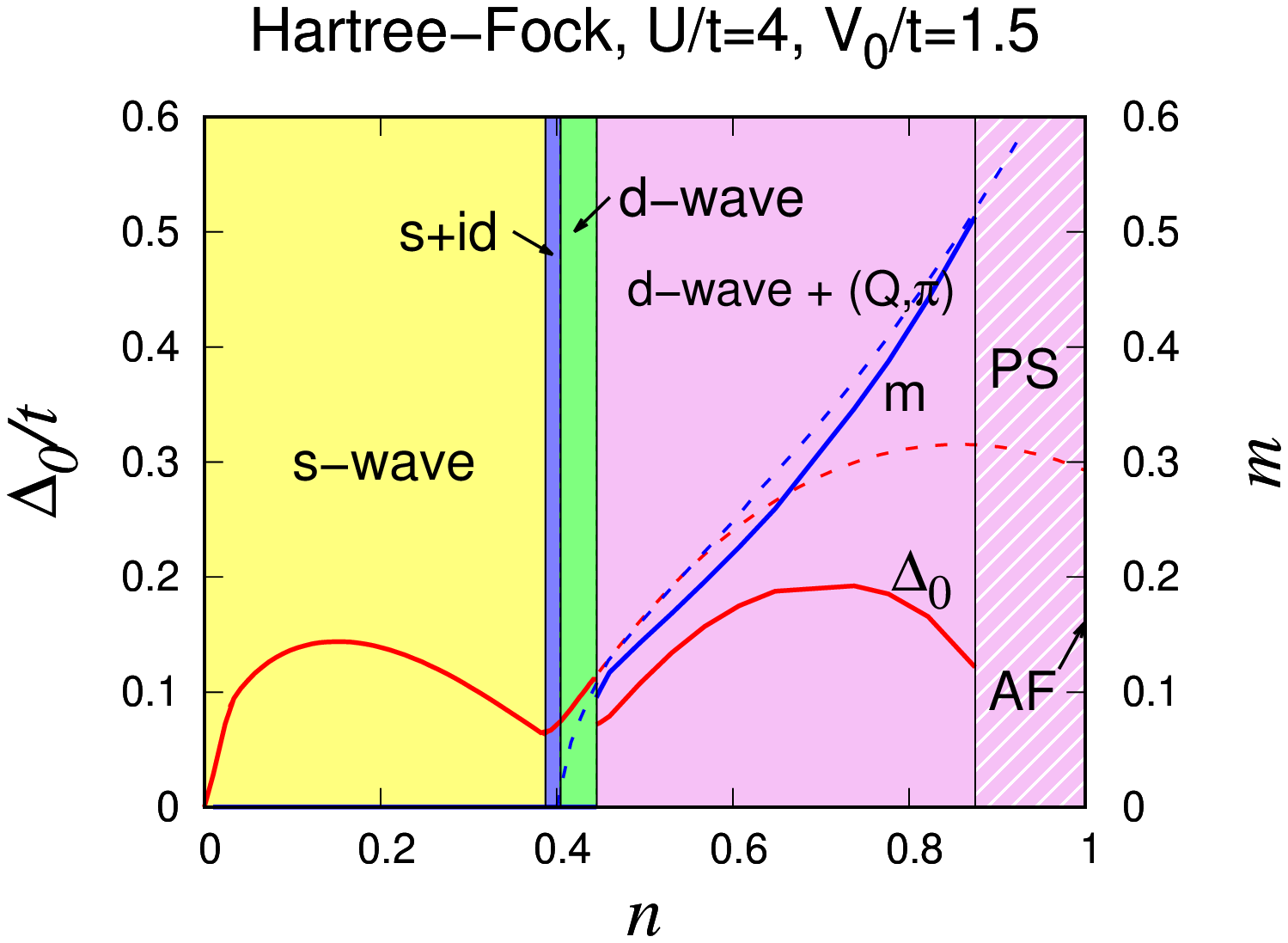}}
	{b)\includegraphics[width=\linewidth]{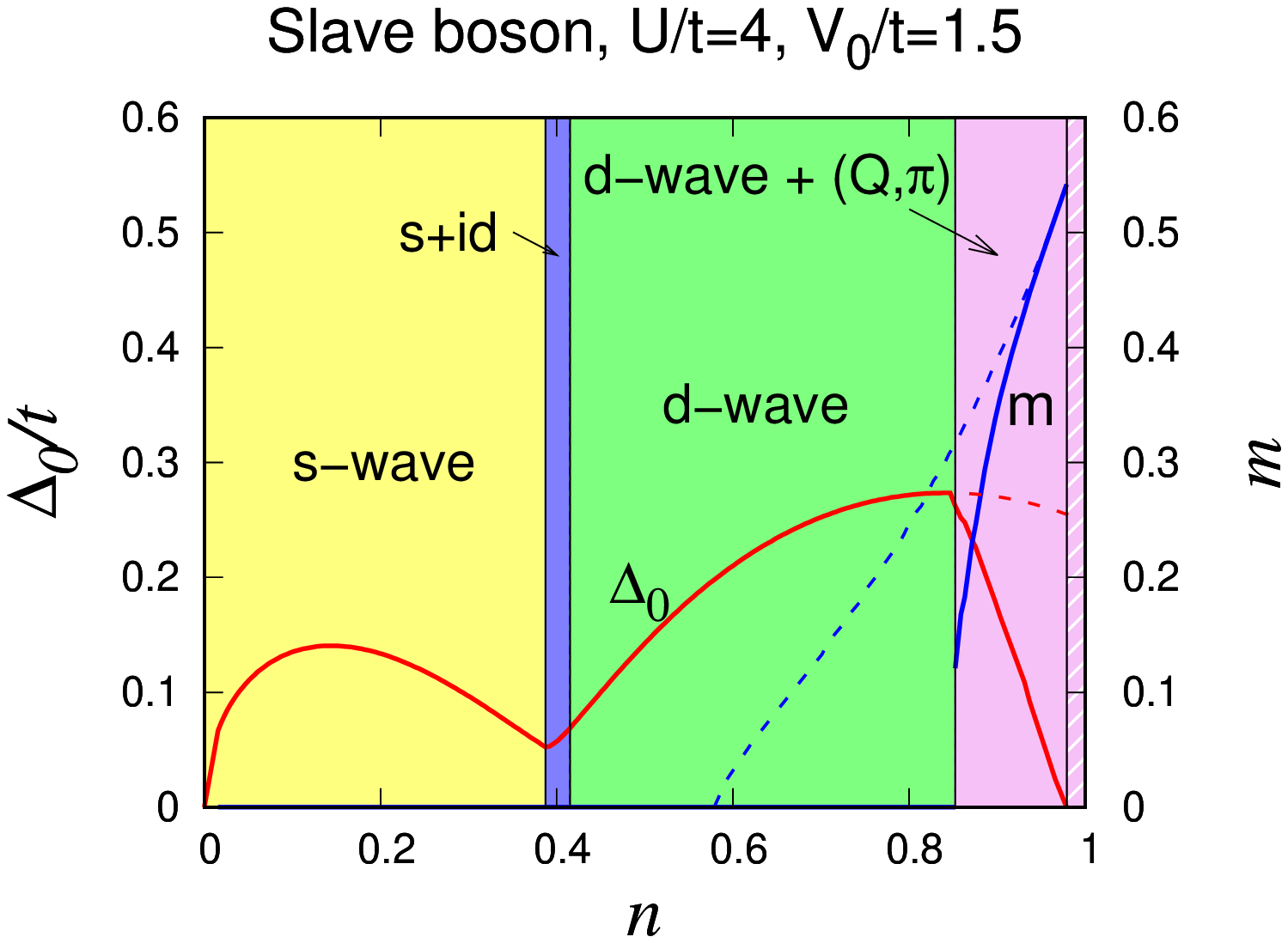}}
	\caption{Dependence of the magnetic $ m $ (blue solid line) and superconducting $ \Delta_0 $ (red solid line) order parameters on the electron concentration $ n $ for $ U = 4t $, $ V_0 = 1 . 5t $ and $ t '= 0 { ,} 2t $.
		The notations $ s $ (yellow), $ d $ (green), and $ s + id $ (blue) correspond to the symmetry of the superconducting order parameter in this region.
		The shaded area denotes the PS region (PS).
		Notations AF and $ (Q, \pi) $ (spiral) correspond to the magnetic order.
		Vertical thin black lines are the boundaries of phase transitions. Thin dashed lines show the dependences of the order parameters in pure magnetic and superconducting systems ($\Delta_0=0$ and $m=0$, correspondingly.
		The symbol "~$ + $~" means the coexistence of orders).
	}
	\label{coex_dep}
\end{figure}
\subsection*{2. $n$-dependence of the order parameters}
The behavior of the magnetic moment $ m $ and the amplitude of the superconducting gap $ \Delta_0 $ illustrated on the figures \ref{coex_dep}a,b, which are constructed corresponding to $ U = 4t $, $ V_0 = 1 . 5t $ and $ t '= 0 . 2t $. 

At low electron concentrations the ground state is the $ s $-wave superconductor. The order parameter behaves non-monotonically with a maximum at $ n \approx0 . 16 $ for both HFA and SBA. At $ n \approx0 . 4 $, a transition to the $ d $-wave superconductor occurs through an intermediate state with the $ s + id $ symmetry of the order parameter. 

At $ n \approx0 . 44 $ for HFA and $n\approx 0.85$ for SBA, a local magnetic moment appears abruptly with an amplitude of $ m \approx0 . 1 $ (this is the first-order phase transition with negligible narrow phase separation area depicted with a single thick red line). Starting from this moment, the superconducting and magnetic orders begin to coexist. In the region of coexistence, the magnetic moment and the amplitude of the superconducting order parameter are smaller compared to pure magnetic and pure superconducting states, the order parameters of which are shown in the figure by dashed lines. Thus, superconductivity and spiral magnetization have a mutually suppressive effect on each other. 

We see the first order phase transition to the insulating AF state, accompanied by a region of PS. In the separation region, a combination of different states is realized: part of the system is insulating AF, while the other part has a spiral magnetic order and is, at the same time, a superconductor. The transition from the superconducting to the dielectric state probably has a percolation nature: the conductivity disappears at the concentration at which the spiral magnetic clusters stop being interconnected. In simple models such a transition occurs at the point at which the fraction of dielectric clusters is equal to $ 1/3 $, which corresponds to the electron concentration~$\approx 0 . 95 $~\cite{Efros}. There is the difference for HFA and SBA methods: HFA diagram have narrow $d$-wave area, but wide coexistence and PS regions, but SBA diagram have wide $d$-wave area and more narrow coexistence and PS regions. The electron correlations provide more favorable conditions for the superconductivity in competition with the spiral magnetism, and have significant influence on $d$-wave superconductivity, less than $s$-wave and $s+$i$d$-wave.

The amplitude of the superconducting gap $\Delta_0 $ behaves non-monotonically in the coexistence region. It grows up to $ \Delta^{max}_0$ and then decreases. Thus, within the framework of our model, it is possible to reproduce the dome-shaped form of the $n$-dependence of the superconducting gap amplitude, which is characteristic of HTSC compounds~\cite{Armitage10, Hosono17}. Traditionally, it is believed that the dome shape is associated with the non-monotonic behavior of the pairing interaction value, which is determined by the nature of the Cooper pairing, for example, unconventional mechanism such as spin fluctuations~\cite{Izyumov99} and other~\cite{Setty22}. Since we do not specify the nature of attraction, and its strength is considered independent of concentration, we show that the mutual influence of the superconducting and magnetic orders can make a sizable contribution into the formation of the dome-shaped dependence $\Delta_0(n)$.

\section{Discussion and conclusions}
We investigate the conditions of coexistence of superconductivity with the intermediate $s+$i$d$ symmetry and spiral magnetic order on a square lattice. A possibility of coexistence and PS between SC and magnetism  is studied by Hartree--Fock and slave boson approaches for $t-U-V$ model. 

The results for HFA and SBA qualitatively similar near $n=1$, but when electron density is far from half filling the correlation effects lead to strong suppression of the variety of magnetic states and the magnetic region width. Hence, the superconductivity becomes more favorable.

It has been shown in \cite{Bulka96} that in the weak-coupling limit the gap in the excitation spectrum, obtained in SBA, is reduced in comparison to that obtained in the HFA. In our investigation $\Delta_0^{HFA}$ is slightly greater than $\Delta_0^{SBA}$ in pure superconducting regime for $d$-wave state,  $\Delta_0^{HFA}\approx \Delta_0^{SBA}$  for $s$-wave and $s+$i$d$-wave states, but in the coexistence mode, the reverse situation is observed: $\Delta_0^{SBA}>\Delta_0^{HFA}$.

In the coexistence regime magnetic moment and $d$-wave superconducting amplitude mutually suppress each other, in agreement with the renormalization group + mean field analysis of the Hubbard model~\cite{Reiss07,Yamase16}, but in our research there is the coexistence with $(Q,\pi)$ spiral magnetic state rather than AF. 

We have analyzed the correlation effects influence on the spiral magnetic and superconducting solutions stability within HFA and SBA comparison, which extends the results obtained in~\cite{Bulka96,Bulka98}.

The Hubbard model phase diagrams were constructed in \cite{Reiss07,Kobayashi10} accounting for $d$-wave superconductivity and commensurate AF magnetic state. The diagrams are similar to ours and in quiet agreement. At the same time, our research takes into account the full set of possible states: $s$-wave, $d$-wave and $s+$i$d$-wave superconductivity, spiral magnetic order and phase transitions between them.

\section{Declaration of Competing Interest}
	The authors declare that they have no known competing financial
	interests or personal relationships that could have appeared to influence
	the work reported in this paper.

\section{Acknowledgements}
The work was carried out within the framework of the state assignment of the Ministry of Science and Higher Education of the Russian Federation (topic \textnumero 121030100005-1).

The authors are grateful to Ph.D. P.A. Igoshev for his contribution to the program code. 
\bibliography{mainbbl}
\bibliographystyle{elsarticle-num}
\end{document}